\documentclass{article}

\usepackage{arxiv}
\usepackage{amsmath}
\usepackage[utf8]{inputenc} 
\usepackage[T1]{fontenc}    
\usepackage{hyperref}       
\usepackage{url}            
\usepackage{booktabs}       
\usepackage{amsfonts}       
\usepackage{nicefrac}       
\usepackage{microtype}      
\usepackage{lipsum}

\title{The counting of Nambu-Goldstone bosons in a non-Hermitian field theory}

\author{Ivan Arraut \\
 The Open University of Hong Kong, 30 Good Shepherd Street, Homantin, Kowloon, Hong Kong, China\\ \\
  \texttt{ivanarraut05@gmail.com} \\
}

\begin{document}
\maketitle

\begin{abstract}
We demonstrate that the number of Nambu-Goldstone bosons is always equal to the number of conserved currents inside the scenario of non-Hermitian field theories with spontaneous symmetry breaking. This eliminates the redundancies which normally appears in Hermitian field theories, specifically when the Lagrangian under analysis violates explicitly the Lorentz symmetry. 
\end{abstract}


\section{Introduction}

When the symmetry of a system, represented by a Lagrangian, is spontaneously broken, gapless particles dubbed Nambu-Goldstone bosons appear. The Nambu-Goldstone theorem then suggests that the number of broken symmetries is equivalent to the number of Nambu-Goldstone bosons \cite{1}. Exceptions to this rule appear when a pair of broken symmetries correspond to only one Nambu-Goldstone boson \cite{2, 3, 4}. Another interesting aspect is that in the standard formulation of the Nambu-Goldstone theorem, the Nambu-Goldstone bosons (being massless particles) are expected to have a linear dispersion relation. However, for the cases where two broken symmetries correspond to a single Nambu-Goldstone boson, it comes out that the dispersion relation is surprisingly quadratic \cite{4}. Nielsen and Chadha formulated an interesting explanation of this phenomena in \cite{2}. Subsequently, Nambu explored the same problem in \cite{5}. Murayama, Watanabe and Brauner formulated some equations explaining the counting rule for the Nambu-Goldstone bosons by extending the arguments of Nielsen and Chadha as well as Nambu's arguments \cite{3}. In \cite{4}, the author demonstrated the generic character of this phenomena which is a natural consequence of the fundamental connection between the Nambu-Goldstone theorem and the Quantum Yang-Baxter equations (QYBE). The formulation of the Nambu-Goldstone theorem in the scenario of the non-Hermitian Field theory (respecting the Parity-time reversal ($PT$) symmetry in order to satisfy unitarity) was done by Alexandre, Ellis, Millingtone and  Seynaeve in \cite{6}. In such a case, the authors demonstrated that for every conserved current, there should be a gapless mode representing the Nambu-Goldstone boson. The conserved current in these situations is not related to a symmetry of the Lagrangian due to the non-Hermitian character of the theory \cite{7}. However, the authors in \cite{6} did not analyze the cases where the Lagrangians under analysis violate explicitly the Lorentz symmetry. Given the fact that this regime (Lorentz violation of symmetry) has showed some redundancies in the ordinary Hermitian theory, in relation with the counting of Nambu-Goldstone bosons, which in some cases is not equal to the number of broken symmetry generators \cite{3, 4}; it is extremely important to analyze the situations where the Lagrangians violate explicitly the Lorentz symmetry inside the scenario of non-Hermitian theories. Note that the redundancies appearing in the case of Hermitian theories, are related to: 1). A special property of the ground state annihilated by a linear combination of symmetry generators. 2). Identities among Noether charge densities \cite{3}. Both of the redundancies are connected with the symmetries of the system which in Hermitian theories are connected to conserved currents. Since $PT$-invariant Hermitian formulations are just a subset of $PT$-invariant non-Hermitian theories, drastic differences between Hermitian and non-Hermitian theories are not expected \cite{Referee}. In this way the redundancies in Hermitian theories appear because we usually take the symmetries of the system as the fundamental objects instead of analyzing the number of independent conserved currents in the system. However, in both situations, Hermitian and non-Hermitian, the number of independent conserved currents decreases when the Lorentz symmetry is explicitly broken by the Lagrangian under analysis. In this paper, by using the K\"ahllen-Lehman spectral representation, we demonstrate that in the case where the Lagrangian obeys the Lorentz symmetry, the dispersion relation of the Nambu-Goldstone bosons is linear and that the number of independent conserved currents is equal to the number of Nambu-Goldstone bosons. On the other hand, for the cases where the Lagrangian violates explicitly the Lorentz symmetry, we demonstrate that the number of conserved currents is reduced with respect to the cases where the Lorentz symmetry is satisfied. The remaining number of independent currents is equal to the number of observed Nambu-Goldstone bosons, this time with quadratic dispersion relation as it should be in this limit. All these results emerge naturally in the non-Hermitian approach proposed in \cite{6, 7}. 
The paper is organized as follows: In Sec. (\ref{S1}), we review the formulation of the Nambu-Goldstone theorem when the Lagrangian under study is non-Hermitian and satisfying the $PT$-symmetry. We then obtain the conditions under which the dispersion relation of the Nambu-Goldstone bosons becomes quadratic by using the K\"allen-Lehman representation. In Sec. (\ref{doublecommutator}), we extend the analysis of the Nambu-Goldstone theorem by expressing the order parameter as a double-commutator and then evaluating the number of independent terms remaining from the final expression. In Sec. (\ref{KLehamannnn}), for a single Lagrangian (system), we analyze the number of independent currents for the relativistic and non-relativistic cases by using the K\"allen-Lehmann representation, demonstrating that the number is reduced by one-half in the non-relativistic limit and it always matches with the number of Nambu-Goldstone bosons. This is the case because a Lagrangian which ordinarily obeys the Lorentz symmetry, when evaluated in the non-relativistic regime, breaks the symmetry explicitly, reducing in this way the number of conserved currents and correspondingly, the number of observed Nambu-Goldstone bosons. Finally in Sec. (\ref{conclusion}), we conclude.              

\section{The Nambu-Goldstone theorem: Non-Hermitian formulation}
\label{S1}

As a starting point, we can assume a transformation for the field $\Phi$ in the following form

\begin{equation}
\Phi\to \Phi+i\epsilon T\Phi,    
\end{equation}
where $T$ is a generator for the transformation. Note that for the case of non-Hermitian Lagrangians we cannot consider symmetry transformations of the Lagrangian, but rather transformations corresponding to a conserved current $j^\mu$ \cite{6, 7}. The corresponding conserved charge is $Q=\int d^3{\bf x}j^0(x)$. The important Mathematical object to be analyzed is the spontaneous symmetry breaking condition

\begin{equation}   \label{eq1}
<\bar{0}\vert[Q, \Phi(x)]\vert0>=iT<\Phi>.    
\end{equation}
Here the left-hand side is a commutator of the Nambu-Goldstone field and the conserved current under analysis. The right-hand side corresponds to the order parameter. Note that for the case of non-Hermitian field theories, the inner products are defined with respect to the $C'PT$ transformation, which is analogous to the standard Charge conjugation-Parity-Time reversal transformation (CPT) defined in the Fock space \cite{6}. Here however, the operator $C'$ does not represent charge conjugation. This is just an operator representing the transformation which guarantees a positive definition of the inner product inside the $PT$ symmetric formulation. Then the bar symbol over the vacuum state projected to the left in eq. ($\ref{eq1}$), is an important distinction to consider inside this formulation. In the same way, the complete set of intermediate states will be defined as $\hat{I}=\sum_n\vert n><\bar{n}\vert$. This means that the covariant version of eq. (\ref{eq1}), can be expressed as

\begin{eqnarray}
<\bar{0}\vert[j^\mu(y), \Phi(x)]\vert0>=\int \frac{d^4p}{(2\pi)^4}e^{-ip(y-x)}\sum_n[(2\pi)^4\delta^4(p_n-p)<\bar{0}\vert j^\mu(0)\vert n><\bar{n}\vert \Phi(0)\vert0>\nonumber\\
-(2\pi)^4\delta^4(p_n+p)<\bar{0}\vert \Phi(0)\vert n><\bar{n}\vert j^\mu(0)\vert0>)].    
\end{eqnarray}
By using the K\"allen-Lehman spectral representation, we can simplify the previous expression as 

\begin{eqnarray}   \label{Kalell}
<\bar{0}\vert[j^\mu(y), \Phi(x)]\vert0>=i\int\frac{d^4p}{(2\pi)^4}e^{-ip(y-x)}2\pi sgn(p_0)p^\mu\rho(p^2).    
\end{eqnarray}
Here $sgn(p_0)$ corresponds to the signature of $p_0=E$, which could be positive ($sgn(p_0)=1$) or negative ($sgn(p_0)=-1$). If we continue the standard procedure, without any surprise, we will obtain a linear dispersion relation. However, if we analyze the previous expressions in detail, we can extract the conditions under which, the dispersion relation becomes quadratic and then we can proceed to analyze the possible consequences of this issue. We will first explore the conditions that the K\"allen-Lehman spectral representation has to satisfy in order to obtain a quadratic dispersion relation. 

\subsection{Conditions for the dispersion relation to be quadratic based on the K\"allen-Lehmann representation}

Following the same arguments illustrated in \cite{4}, we can obtain the general conditions for the dispersion of the Nambu-Goldstone bosons to be quadratic. The K\"allen-Lehmann representation by itself cannot tell us anything about the counting of Nambu-Goldstone bosons, at least in its standard form. The same representation can however, help us to understand under which circumstances we can expect the quadratic dispersion relation. The starting point is to rewrite the result (\ref{Kalell}) in a more explicit form as

\begin{eqnarray}   \label{This onela}
<\bar{0}\vert[j^\mu(y), \Phi(x)]\vert0>=\int\frac{d^4p}{(2\pi)^4}\left(2\pi i\Theta(p_0)p^\mu\rho(p^2)e^{-ip_n(y-x)}-2\pi i\Theta(-p_0)p^\mu\bar{\rho}(p^2)e^{ip_n(y-x)}\right).    
\end{eqnarray}
Here $\Theta(x)$ corresponds to the Heaviside function which is non-trivial when $x\geq0$ such that $\Theta(x\geq0)=1$.    
This previous expression can be factorized as

\begin{equation}   \label{This onela2}
<\bar{0}\vert[j^\mu(y), \Phi(x)]\vert0>=\int\frac{d^4p}{(2\pi)^4}2\pi i\Theta(p_0)e^{-ip^0(y^0-x^0)}\left(p^\mu\rho(p^2)e^{i\vec{p}_n\cdot(\vec{y}-\vec{x})}-\bar{p}^\mu\bar{\rho}(p^2)e^{-i\vec{ p}_n\cdot(\vec{y}-\vec{x})}\right).    
\end{equation}
Here $p^\mu=(E, \vec{p})$ and $\bar{p}^\mu=(-E, \vec{p})$. Then in order to analyze the previous expression properly, it is convenient to expand its product with respect to a four-vector $n^\mu=(1, \vec{u})$, where $\vec{u}$ is a unit space-like vector. Considering

\begin{equation}
n_\mu p^\mu=E-\vec{u}\cdot\vec{p},\;\;\;\;\; n_\mu \bar{p}^\mu=-E-\vec{u}\cdot\vec{p}, 
\end{equation}
where we have defined the form $n_\mu=(1, -\vec{u})$, which  multiplied by eq. (\ref{This onela2}), gives 

\begin{equation}   \label{This onela2again}
n_\mu<\bar{0}\vert[j^\mu(y), \Phi(x)]\vert0>=\int\frac{d^4p}{(2\pi)^4}4\pi i\Theta(p_0)\rho(p^2)e^{-ip^0(y^0-x^0)}\left(E cos(\vec{ p}_n\cdot(\vec{y}-\vec{x}))-i\vec{u}\cdot\vec{p}sin(\vec{p}_n\cdot(\vec{ y}-\vec{x}))\right).    
\end{equation}
Here we have considered that $\rho(p^2)=\bar{\rho}(p^2)$ in agreement with the arguments of causality illustrated in \cite{4}. Note that in the non-relativistic case $E\gg\vert\vec{p}\vert$, from the previous expression we obtain

\begin{equation}   \label{This onela22}
n_\mu<\bar{0}\vert[j^\mu(y), \Phi(x)]\vert0>\approx\int\frac{d^4p}{(2\pi)^4}4\pi i\Theta(p_0)\rho(p^2)e^{-ip^0(y^0-x^0)}E cos(\vec{ p}_n\cdot(\vec{y}-\vec{x})).    
\end{equation}
Here we have neglected the contributions coming from the term proportional to $\vec{u}\cdot\vec{p}$ in eq. (\ref{This onela2again}). Since the cosine function expands as a quadratic function at the lowest order in its argument, then with the result (\ref{This onela22}), we have demonstrated that $E\approx\vert\vec{p}\vert^2$ for this case. This can be observed from eq. (\ref{This onela22}) because the energy in addition expands linearly at the lowest order. Then at the moment of evaluating the limit $E\to0$, simultaneously with the limit $\vec{p}\to0$, we observe that the term depending on $E$ in eq. (\ref{This onela22}) goes to zero linearly and the term depending on $\vec{p}$ goes to zero quadratically. This method of evaluating the vanishing limits for the frequency (energy) and the momentum in order to understand the dispersion relations, are the same methods used traditionally in \cite{81}, with the difference that this time we are using the K\"allen-Lehmann representation instead of the standard commutator expansion. Since a physical system in the non-relativistic regime violates explicitly the Lorentz symmetry, the result showed here is consistent with the claims suggesting that when the Lorentz symmetry is violated explicitly by the Lagrangian, the Nambu-Goldstone bosons have a quadratic dispersion relation \cite{1}. Note that our derivation is not concerned with the exact functional dependence of $\rho(p^2)$. On the other hand, in the relativistic limit, we have $E\approx \vert\vec{p}\vert$, and from the result (\ref{This onela2again}), we get

\begin{equation}   \label{This onela223}
n_\mu<\bar{0}\vert[j^\mu(y), \Phi(x)]\vert0>\approx \int\frac{d^4p}{(2\pi)^4}4\pi i\Theta(p_0)\rho(p^2)Ee^{-ip^0(y^0-x^0)}\left(cos(\vec{p}_n\cdot(\vec{y}-\vec{x}))-i(cos\alpha) sin(\vec{p}_n\cdot(\vec{y}-\vec{x}))\right). 
\end{equation}
Here we have used the result $\vec{u}\cdot\vec{p}=\vert\vec{u}\vert\vert\vec{p}\vert cos\alpha$, with $\alpha$ being the angle between $\vec{u}$ and $\vec{p}$ and the corresponding vector norms satisfying the conditions $\vert\vec{u}\vert=1$ and $\vert\vec{p}\vert\approx E$. At the lowest order in the momentum expansion, we have $cos(\vec{p}_n\cdot(\vec{y}-\vec{x}))\approx 1$ and $sin(\vec{p}_n\cdot(\vec{y}-\vec{x}))\approx \vec{p}_n\cdot(\vec{y}-\vec{x})$, and then eq. (\ref{This onela223}) becomes

\begin{equation}   \label{Rapunzel}
n_\mu<\bar{0}\vert[j^\mu(y), \Phi(x)]\vert0>\approx \int\frac{d^4p}{(2\pi)^4}4\pi i\Theta(p_0)\rho(p^2)Ee^{-ip^0(y^0-x^0)}(1-i\vec{p}_n\cdot(\vec{y}-\vec{x})(cos\alpha)). 
\end{equation}

In this previous expression, when we evaluate the limits $E\to0$ and $\vec{p}\to0$ simultaneously, it is clear that the frequency (energy) goes to zero linearly and the momentum also goes to zero linearly. Since the regime under evaluation is relativistic, the system here respects the Lorentz symmetry explicitly. Then we have a dispersion relation of the form $E\approx \vert\vec{p}\vert$ for the massless Nambu-Goldstone bosons as it is expected in this limit where the Lorentz symmetry is satisfied explicitly by the Lagrangian. Note that here nothing can be concluded about the relation between the number of Nambu-Goldstone bosons and the conserved currents. Something important to remark is that in order to understand the way how the Nambu-Goldstone disperse, we must focus on the behavior of the phases $e^{\pm ipy}$ and the way how they factorize at the end. This was the key part in the analysis done in \cite{4}.     

\section{Double-commutator structure: The standard case based on broken generators}   \label{doublecommutator}

In order to find explicitly the connection between the number of Nambu-Goldstone bosons and the number of independent currents, we have to develop the expression (\ref{This onela}) further. Before doing this, in the ordinary case, it was assumed that the broken symmetries of the system obeyed the Lie-algebra structure. This suggests that the following expression should be satisfied

\begin{equation}
[Q_p, Q_l]=C_{plc}Q_c,    
\end{equation}
with $C_{plc}$ denoting the structure constant. By extending the arguments to the case of conserved currents, we can convert the result (\ref{This onela2}) into the following expression

\begin{equation}
<\bar{0}\vert[Q_c^\gamma(w), \Phi_b(x)]\vert0>\simeq<\bar{0}\vert [[Q_p^\mu(y), Q_l^\nu(z)], \Phi_b(x)]\vert0>,    
\end{equation}
which is a double-commutator with four terms after expanding the expression. If we evaluate explicitly the double commutator, then we obtain

\begin{equation}
<\bar{0}\vert Q_p^\mu(y) Q_l^\nu(z)\Phi_b(x)\vert0>-<\bar{0}\vert Q_l^\nu(z) Q_p^\mu(y)\Phi_b(x)\vert0>+<\bar{0}\vert \Phi_b(x)Q_l^\nu(z) Q_p^\mu(y)\vert0>-<\bar{0}\vert\Phi_b(x)Q_p^\mu(y) Q_l^\nu(z)\vert0>\neq0. 
\end{equation}
We can introduce now a pair of complete set of intermediate states, which for the present non-Hermitian field theory are equivalent to $\hat{I}=\sum_{m}\vert m><\bar{m}\vert$ and $\hat{I}=\sum_{n}\vert n><\bar{n}\vert$. Then we obtain

\begin{eqnarray}   \label{tripleX2}
\sum_{m, n}\left(<\bar{0}\vert Q_p^\mu(y)\vert n><\bar{n}\vert Q_l^\nu(z)\vert m><\bar{m}\vert\Phi_b(x)\vert0>
-<\bar{0}\vert Q_l^\nu(z) \vert m><\bar{m}\vert Q_p^\mu(y)\vert n><\bar{n}\vert\Phi_b(x)\vert0>\right)\nonumber\\
+\sum_{m, n}\left(<\bar{0}\vert \Phi_b(x)\vert m><\bar{m}\vert Q_l^\nu(z)\vert n><\bar{n}\vert Q_p^\mu(y)\vert0>-<\bar{0}\vert\Phi_b(x)\vert n><\bar{n}\vert Q_p^\mu(y)\vert m><\bar{m}\vert Q_l^\nu(z)\vert0>\neq0\right).
\end{eqnarray}
Here the sum is applied to all the terms. Under the space-time translational invariance assumption, then we get the more explicit expression

\begin{eqnarray}   \label{tripleX}
\int d^4p d^4\tilde{p}e^{-ip(y-z)}e^{-i\tilde{p}(z-x)}\sum_{m, n}\delta^{(4)}(p_n-p)\delta^{(4)}(p_m-\tilde{p})<\bar{0}\vert Q_p^\mu(0)\vert n><\bar{n}\vert Q_l^\nu(0)\vert m><\bar{m}\vert\Phi_b(0)\vert0>\nonumber\\
-\int d^4p d^4\tilde{p}e^{-i\tilde{p}(z-y)}e^{-ip(y-x)}\sum_{m, n}\delta^{(4)}(p_n-p)\delta^{(4)}(p_m-\tilde{p})<\bar{0}\vert Q_l^\nu(0) \vert m><\bar{m}\vert Q_p^\mu(0)\vert n><\bar{n}\vert\Phi_b(0)\vert0>\nonumber\\
+\int d^4p d^4\tilde{p}e^{-i\tilde{p}(x-z)}e^{-ip(z-y)}\sum_{m, n}\delta^{(4)}(p_n-p)\delta^{(4)}(p_m-\tilde{p})<\bar{0}\vert \Phi_b(0)\vert m><\bar{m}\vert Q_l^\nu(0)\vert n><\bar{n}\vert Q_p^\mu(0)\vert0>\nonumber\\
-\int d^4p d^4\tilde{p}e^{-ip(x-y)}e^{-i\tilde{p}(y-z)}\sum_{m, n}\delta^{(4)}(p_n-p)\delta^{(4)}(p_m-\tilde{p})<\bar{0}\vert\Phi_b(0)\vert n><\bar{n}\vert Q_p^\mu(0)\vert m><\bar{m}\vert Q_l^\nu(0)\vert0>\neq0.
\end{eqnarray}
Previous arguments suggests that the first and the fourth terms of the previous expression represent the same sequence of events \cite{4} after summing over the degenerate vacuum. The same applies for the second and the third term in the previous equation. Such equality corresponds to two pairs of Quantum Yang-Baxter Equations (QYBE). Although the formulation showed here is valid in the ordinary sense, it requires some modifications when we deal with non-Hermitian formulations. In what follows, we will apply the K\"allen-Lehmann spectral representation. In such a case, since the Wick Theorem forbids non-trivial triple products at the moment of evaluating propagators \cite{8}, we have to formulate an equivalent way to represent the triple products appearing previously in the language of the K\"allen-Lehmann representation. This implies the analysis of the product of a pair of two-point functions or equivalently, exploring the product of two propagators. Note that in this section we developed the standard method which appear in any text-book, nothing different to that. The coming section comes with the new ingredients to be analyzed. 

\section{The K\"allen-Lehmann representation of the triple product: The number of independent currents and Nambu-Goldstone bosons}   \label{KLehamannnn}

Since the ordinary concept of propagator does not match with triple-products, the K\"allen-Lehmann representation adapted to the case of the analysis of a pair of conserved currents inside the scenario of spontaneous symmetry breaking, must be expressed as the product of two propagators. Then, instead of considering the product of three matrices as in the equation (\ref{tripleX}), we evaluate the following product of propagators

\begin{equation}   \label{propa}
\left(<\bar{0}\vert[\Phi(x), j_p^\mu(y)]\vert0>\right)\left([<\bar{0}\vert [j_l^\nu(z), \Phi(x)]\vert0>\right)=D^\mu_{l\;F}(x-y)D^\nu_{p\;F}(z-x).   
\end{equation}
Here $D^\mu_F(x)$ and $D^\nu_F(x)$ correspond to the Feynman propagators. These mathematical objects are matrices with the entries corresponding to the spacetime labels $\mu$ and $\nu$. The same matrices contain the additional labels $l$ and $p$, which correspond to the different conserved currents inside the system. Note that if we use the Wick's theorem \cite{8}, we must consider all the possible contractions between the fields, which would correspond to the product of two propagators. However, since eq. (\ref{propa}) considers the time-ordered product of two currents with the same Nambu-Goldstone field $\Phi(x)$ (repeated twice), then the following propagator is trivial 

\begin{equation}
\left(<\bar{0}\vert[\Phi(x),\Phi(x)]\vert0>\right)=0.    
\end{equation}
The other possible contraction (or propagator) from the product of the four fields in eq. (\ref{propa}) just corresponds to the same result defined in eq. (\ref{propa}). Then if we apply the Wick's theorem to the product $\Phi(x) j_p^\mu(y)j_l^\nu(z)\Phi(x)$, the only non-trivial result corresponds to the product of the two propagators represented in eq. (\ref{propa}). By expressing the result in terms of the K\"allen-Lehmann representation, from eq. (\ref{propa}) we obtain 

\begin{eqnarray}   \label{tobeshown}
\left(<\bar{0}\vert[\Phi(x), j_p^\mu(y)]\vert0>\right)\left([<\bar{0}\vert [j_l^\nu(z), \Phi(x)]\vert0>\right)=-\int\frac{d^4p}{(2\pi)^4}\frac{d^4\tilde{p}}{(2\pi)^4}4\pi^2\Theta(p_0)\Theta(\tilde{p}_0)\rho(p^2)\rho(\tilde{p}^2)e^{-i\tilde{p}^0(z^0-x^0)}e^{-ip^0(x^0-y^0)}\nonumber\\
\times\left(p^\mu\tilde{p}^\nu e^{i\vec{\tilde{p}}_m\cdot(\vec{z}-\vec{x})}e^{i\vec{ p}_n\cdot(\vec{x}-\vec{y})}+\bar{p}^\mu\bar{\tilde{p}}^\nu e^{-i\vec{\tilde{p}}_m\cdot(\vec{z}-\vec{x})}e^{-i\vec{ p}_n\cdot(\vec{x}-\vec{y})}-p^\mu\bar{\tilde{p}}^\nu  e^{-i\vec{\tilde{p}}_m\cdot(\vec{z}-\vec{x})}e^{i\vec{ p}_n\cdot(\vec{x}-\vec{y})}-\bar{p}^\mu\tilde{p}^\nu e^{i\vec{\tilde{p}}_m\cdot(\vec{z}-\vec{x})}e^{-i\vec{ p}_n\cdot(\vec{x}-\vec{y})} \right).
\end{eqnarray}
Here again $p^\mu=(E, \vec{p})$, $\bar{p}^\mu=(-E, \vec{p})$, with analogous definitions for $\tilde{p}^\nu$ and $\bar{\tilde{p}}^\nu$. Here we define the matrix elements in eq. (\ref{tobeshown}) as

\begin{eqnarray}   \label{justwatch}
p^\mu\tilde{p}^\nu=
\begin{pmatrix} 
E\tilde{E} & E\vec{\tilde{p}}^T \\
\vec{p}\tilde{E} & \vec{p}\vec{\tilde{p}}^T 
\end{pmatrix},\;\;\;\;\;
\bar{p}^\mu\bar{\tilde{p}}^\nu=
\begin{pmatrix} 
E\tilde{E} & -E\vec{\tilde{p}}^T \\
-\vec{{p}}\tilde{E} & \vec{{p}}\vec{\tilde{p}}^T 
\end{pmatrix},\;\;\;\;\;
p^\mu\bar{\tilde{p}}^\nu=
\begin{pmatrix} 
-E\tilde{E} & E\vec{\tilde{p}}^T \\
-\vec{p}\tilde{E} & \vec{p}\vec{\tilde{p}}^T 
\end{pmatrix},\;\;\;\;\;
\bar{p}^\mu\tilde{p}^\nu=
\begin{pmatrix} 
-E\tilde{E} & -E\vec{\tilde{p}}^T \\
\vec{p}\tilde{E} & \vec{p}\vec\tilde{p}^T
\end{pmatrix}.
\end{eqnarray}
If we introduce these matrix definitions inside eq. (\ref{tobeshown}), we can then analyze the two regimes, namely, the relativistic regime (respecting Lorentz symmetry) and the non-relativistic one (explicit violation of Lorentz symmetry). Note that the two columns appearing in the previous matrices represent in reality four columns corresponding to the 4-components of the currents, they are: A temporal component and three spatial components. The three spatial components are redundant and they can be reduced to one if we select a basis with the unit vector parallel to the momentum vector. Then we can reduce the number of independent columns to two, which corresponds to the number of Nambu-Goldstone bosons appearing in the system, consistent with the tensorial products $p^\mu\tilde{p}^\nu$, $\bar{p}^\mu\bar{\tilde{p}}^\nu$, $p^\mu\bar{\tilde{p}}^\nu$ and $\bar{p}^\mu\tilde{p}^\nu$. Another way to visualize that the matrices can be reduced to two columns is by noticing that what we are doing in eq. (\ref{justwatch}) is just a tensorial product of two vectors, each one corresponding to two different propagators as can be seen from eq. (\ref{propa}). Then at the end we will only have two independent columns for the matrices (\ref{justwatch}), no matter how many spatial components the currents have. Then we can safely reduce the $4\times4$ to a $2\times2$ matrix. 

\subsection{Lorentz Symmetry violating case: Non-relativistic regime}

For the non-relativistic regime, $E\gg\vert\vec{p}\vert$, then the matrices defined in eq. (\ref{justwatch}), are reduced to only one independent column. In other words, in this regime, all the matrices defined in eq. (\ref{justwatch}) are reduced to 

\begin{equation}   \label{justwatch2}
p^\mu\tilde{p}^\nu\approx \bar{p}^\mu\bar{\tilde{p}}^\nu\approx p^\mu\bar{\tilde{p}}^\nu\approx\bar{p}^\mu\tilde{p}^\nu\approx   
\begin{pmatrix} 
E\tilde{E} & 0 \\
0 & 0 
\end{pmatrix}.
\end{equation}
Here we have neglected any term proportional to $\vec{p}$ or $\vec{\tilde{p}}$ in eq. (\ref{justwatch}). Then a matrix with multiple entries is now reduced to a single scalar component proportional to $E\tilde{E}$. If we introduce this result in eq. (\ref{tobeshown}), then we get

\begin{eqnarray}   \label{almost}
\left(<\bar{0}\vert[\Phi(x), j_p^\mu(y)]\vert0>\right)\left([<\bar{0}\vert [j_l^\nu(z), \Phi(x)]\vert0>\right)\approx -\int\frac{d^4p}{(2\pi)^4}\frac{d^4\tilde{p}}{(2\pi)^4}16\pi^2E\tilde{E}\Theta(p_0)\Theta(\tilde{p}_0)\rho(p^2)\rho(\tilde{p}^2)e^{-i\tilde{p}^0(z^0-x^0)}\nonumber\\
\times
e^{-ip^0(x^0-y^0)}cos(\vec{\tilde{p}}_m\cdot(\vec{z}-\vec{x}))cos(\vec{ p}_n\cdot(\vec{x}-\vec{y})).
\end{eqnarray}
Then what was initially a $4\times4$ matrix (considering one temporal and three spatial components) with two independent columns, becomes a single component. This means that in the non-relativistic regime (regime where the Lorentz symmetry is violated explicitly), the number of non-redundant conserved currents is reduced by one half. This in addition means that the number of Nambu-Goldstone bosons is also reduced correspondingly. Before we have demonstrated in eq. (\ref{This onela22}) that the dispersion relation of the Nambu-Goldstone bosons is quadratic in these cases. Since the the pair of independent currents are reduced to a single one in this case, then we can safely assume that $p^0\approx \tilde{p}^0$ and $\vec{ p}_n\approx\vec{\tilde{p}}_m$ in eq. (\ref{almost}).

\subsection{Lorentz symmetry case: Relativistic regime}

For the relativistic regime, we have that the condition $E \approx \vert\vec{p}\vert$ is satisfied. Then eq. (\ref{tobeshown}) becomes

\begin{eqnarray}
\left(<\bar{0}\vert[\Phi(x), j_p^\mu(y)]\vert0>\right)\left([<\bar{0}\vert [j_l^\nu(z), \Phi(x)]\vert0>\right)=-\int\frac{d^4p}{(2\pi)^4}\frac{d^4\tilde{p}}{(2\pi)^4}4\pi^2\Theta(p_0)\Theta(\tilde{p}_0)\rho(p^2)\rho(\tilde{p}^2)e^{-i\tilde{p}^0(z^0-x^0)}\nonumber\\
\times e^{-ip^0(x^0-y^0)}M,
\end{eqnarray}
where $M$ is a complex matrix defined as

\begin{equation}
M=4E\tilde{E}
\begin{pmatrix}
 cos(\vec{\tilde{p}}_m\cdot(\vec{z}-\vec{x}))cos(\vec{ p}_n\cdot(\vec{x}-\vec{y}))&i sin(\vec{\tilde{p}}_m\cdot(\vec{z}-\vec{x}))cos(\vec{ p}_n\cdot(\vec{x}-\vec{y}))\\
 -sin(\vec{\tilde{p}}_m\cdot(\vec{z}-\vec{x}))sin(\vec{ p}_n\cdot(\vec{x}-\vec{y}))&i cos(\vec{\tilde{p}}_m\cdot(\vec{z}-\vec{x}))sin(\vec{ p}_n\cdot(\vec{x}-\vec{y}))
\end{pmatrix}.
\end{equation}
The fact that the matrix preserves the two linearly independent vectors (columns), means that the two currents under analysis, are relevant in this regime. Then the number of Nambu-Goldstone bosons is not reduced in these situations because in general $E\neq \tilde{E}$ as well as $\vec{ p}_n\neq\vec{\tilde{p}}_m$. Already in eq. (\ref{This onela223}) we have demonstrated that in this regime the dispersion relation is linear. 

\section{Conclusions}   \label{conclusion}

In this paper we have demonstrated that the non-Hermitian formulation of Quantum-Field theory avoids naturally the redundancies in the counting of Nambu-Goldstone (NG) bosons by connecting the number of NG bosons directly with the number of independent conserved currents in the system. Then while in the relativistic case, where the Lorentz symmetry is respected explicitly, we have $n$ Nambu-Goldstone bosons with linear dispersion relation, connected to the existence of $n$ independent conserved currents; in the non-relativistic case, where the Lorentz symmetry is violated explicitly, the number of independent conserved currents as well as the number of Nambu-Goldstone bosons is reduced by a half to $n/2$. The dispersion relation of the Nambu-Goldstone bosons is quadratic as it is expected in the cases of explicit violation of the Lorentz symmetry. Note that in this paper, when we talk about relativistic regime, we are talking about Lagrangians respecting explicitly the Lorentz invariance. On the other hand, the non-relativistic regime corresponds to the situations where the Lorentz invariance of the Lagrangian describing the dynamics of the system is explicitly broken \cite{3, 4}. In both situations, the Nambu-Goldstone bosons, being massless behave in a relativistic fashion, although the dispersion relation changes from one regime to the other. The information about the dispersion relations, is contained in the phases appearing in the obtained expressions.  
Within the two regimes analyzed in this paper, the spontaneous symmetry breaking does not correspond to the Lorentz symmetry breaking but rather to the spontaneous breaking of internal symmetries. In this paper we have used the K\"ahllen-Lehmann spectral representation for our analysis by extending the arguments illustrated in \cite{6, 7}. Note that in \cite{6, 7}, the authors did not analyze the situations where the Lorentz symmetry is violated explicitly (non-relativistic regime in this paper) in order to understand possible emerging redundancies in the counting rules for Nambu-Golstone bosons \cite{3, 4}. It is important to remark in addition that for the case of Hermitian theories, the results obtained for the dispersion relations of the Nambu-Goldstone bosons are the same as those obtained in the non-Hermitian case. However, in the Hermitian regime, the relation between the number of Nambu-Goldstone bosons and the number of broken symmetries is not fixed when we analyze the different regimes (Lorentz violating regimes or regimes respecting the Lorentz symmetry). This means that while the standard Hermitian formulation of Quantum Field Theory might shows some redundancies in the counting rules for Nambu-Goldstone bosons, these ambig\"uities disappear for the case of non-Hermitian field theories. These important aspects remark the importance of understanding in deep detail the non-Hermitian formulations as those proposed in \cite{6, 7}. A huge deviation in the physics around the NG bosons is not expected when we compare the Hermitian and the non-Hermitian formulations \cite{Referee}. In this way, the source of the redundancies in the Hermitian theories is the fact that the symmetries of the system are taken as fundamental in such a case. However, what should be really considered fundamental are the conserved currents emerging within the system which are also reduced in the case of Hermitian theories \cite{3}. In the case of Hermitian theories, it has been proved for example that some symmetries become redundant when the Lorentz symmetry is violated \cite{3}. Since the non-Hermitian formulation is based on conserved currents, completely disconnected from symmetries, these ambig\"uities related to the existence of symmetries in the system are not perceived. Thus even in the Hermitian case, we have to consider the conserved currents as more fundamental than the symmetries of the system. However, it is still precious to make connections between symmetries and conserved currents.   

\bibliographystyle{unsrt}  


\end{document}